\documentclass[prb,nofootinbib,twocolumn,superscriptaddress]{revtex4} 

\usepackage{graphicx}
\usepackage{dcolumn}
\usepackage{bm}
\usepackage{threeparttable}
\usepackage{times}
\usepackage{mathptmx}
\usepackage{lscape}
\usepackage{natbib}
\usepackage{amsmath}
\usepackage{amssymb}
\usepackage{braket}

\def\degree{\kern-.2em\r{}\kern-.3em}

\begin{document}


\title{ Structure of Non-Solid Matter in Equilibrium State under $NVT$ ensemble: \\ New Insight from Spatial Constraint}

\author{Koretaka Yuge}
\affiliation{
Department of Materials Science and Engineering,  Kyoto University, Sakyo, Kyoto 606-8501, Japan\\
}%

\author{Makoto Murata}
\affiliation{
Department of Materials Science and Engineering,  Kyoto University, Sakyo, Kyoto 606-8501, Japan\\
}%

\begin{abstract}
When non-solid matter (e.g., liquids or gas) is under constant volume $V$ and density $\rho$ (e.g., in rigid box), spatial positions for their constituents are restricted by these conditions. 
We recently focus on the role of constraint in classical statistical thermodynamics, and find how spatial constraint connects with equilibrium properties for crystalline solids, which has not been clarified so far. The present study extend the idea to non-solid matter under $NVT$ ensemble in classical systems. We provide explicit representation of canonical average of radial distribution function in terms of spatial constraint, which can be well characterized by a single special microscopic state on configuration space called "projection state" for non-solid matter. 
We demonstrate that the special microscopic state can be numerically constructed for a finite number of particles. 

\end{abstract}


\maketitle
\section{Introduction}
In classical systems where total energy is the sum of kinetic and potential energy for constituents, expectation value of structure along chosen coordination $q_r$ in equilibrium state (we prepara complese set of basis functions corresponding to the coordination $\left\{q_1,\ldots, q_g \right\}$) is generally given by canonical average, 
\begin{eqnarray}
\label{eq:qcan}
Q_r\left(T\right) = Z^{-1} \sum_{i} q_r^{\left(i\right)} \exp\left(-\beta E^{\left(i\right)}\right),
\end{eqnarray}
where summation is taken over all possible microscopic states on phase space, and $\beta=\left(k_{\textrm{B}}T\right)^{-1}$.
Since energy for individual microscopic state should depend on constituents and their multibody interactions, a set of microscopic state to dominantly determine equilibrium properties cannot be known \textit{a priori} until information about temperature and interactions are given.
Despite this fact, we recently develop theoretical approach, where a specially selected microscopic state, that can be known \textit{a priori} without any information about energy or temperature, can well characterize the equilibrium properties. This result is naturally derived by clarifying how equilibrium properties connect with spatial constraint on the system, which can tell us the 
important microscopic states describing equilibrium states, based on information only about the spatial constraint. 
Consequently, Eq.~(\ref{eq:qcan})  can be greatly simplified to
\begin{eqnarray}
\label{eq:qe}
Q_r\left(T\right) \simeq \Braket{q_r}_1 - \sqrt{\frac{\pi}{2}}\Braket{q_r}_2 \frac{U_r^{+} - U_0}{k_{\textrm{B}}T},
\end{eqnarray}
where $\Braket{\cdot}_1$ and $\Braket{\cdot}_2$ respectively denotes taking average and standard deviation over microscopic states on configuration space without weight of Boltzmann factor, and $U_r^+$ and $U_0$ denotes potential energy of special microscopic states (for the former, we call "projection state", and for the latter, we call "mean state" hereinafter), whose microscopic structures can be determined from information about spatial constraint. 
Validity and applicability of proposed equation of Eq.~(\ref{eq:qe}) has been confirmed for crystalline solids in our previous studies. 
Although Eq.~(\ref{eq:qe}) can be applied for any classical systems, application to non-solid matter including liquid and gas is non-trivial. 
This is because even under the same volume and density, their constituents themselves can have different effective size, while straightforward extention of our theoretical approach in 
crystalline solids to non-solid matter does not explicitly include this size effect. Furthermore, whether or not the special microscopic states of projection and mean state can be practically constructed for non-solid matter is not confirmed so far. 
Therefore, in order to confirm the validy and applicability of our theoretical approach, the present study first provides explicit representation of canonical average of radial distribution function, and then, applies the representation to non-solid state whose interaction is given by Lenard-Jones potential with various potential parameters. 
We find that similary to crystalline solids, special microscopic state for non-solid state can be numerically constructed for finite-size system, and canonical average of radial distribution function can be reasonablly predicted for multiple particle size. 

\section{Derivation and Concept}
Since Eq.~(\ref{eq:qe}) can be applied to any coordination that can be linear transformed from for instance, coordination in generalize Ising model (GIM) used in our previous studies. 
Therefore, to describe microscopic structures, we here derive the explicit expression of Eq.~(\ref{eq:qe}) for the case of well-known radial distribution function (RDF) by clarifying relationship between GIM description and RDF. Consider the system in liquid phase is in constant volume $V$ and in constant number density $\rho$. Let us focus on a single particle $i$ for a certain microscopic state, and define the number of particles, whose distance from particle $i$ is between $r$ and $r + dr$, as $n_{i}\left(r\right)$. Then the radial distribution function is given by
\begin{eqnarray}
\label{eq:g}
g\left(r\right) = \frac{\overline{n\left(r\right)}}{4\pi r^{2}dr\rho},
\end{eqnarray}
where $\overline{n\left(r\right)}$ denotes average of $n_{i}\left(r\right)$ over all constituent particles. Therefore the radial distribution function is a measure of average density in given distance $r$ in terms of that in ideally random mixture. Practically, expectectation value of  the radial distribution function, $\left<g\left(r\right)\right>_{Z}$, is required to determine atomic structure of the liquid phase. The goal of the present study is thus to first derive a representation of $\left<g\left(r\right)\right>_{Z}$ based on our theory,\cite{lsi,emrs, rmx} which would provide new insight into how radial distribution in liquid phase is ruled by distance $r$ and would enable efficient estimation of $\left<g\left(r\right)\right>_{Z}$ independent of the choice of constituent elements. 

Let us divide the three dimensional system into equivalent and sufficiently small cubes, where each cube has a length of $R$ on one side. When we interpret vertex of the cubes as lattice points, the number of lattice points in the system and number of particles are respectively given by $V/R^{3}$ and $\rho V$. We define the composition $x$ as
\begin{eqnarray}
x = \cfrac{\rho V}{V/R^{3}},
\end{eqnarray}
which represents fraction of the number of particles with respect to the number of lattice points. We then introduce spin variable $\sigma_{k}$, where $\sigma_{k}=+1$ denote occupation of lattice point $k$ by particle and $\sigma_{k}=-1$ is a vacant site. With this definition, number of pair sites in distance between $r$ and $r+dr$ is given by
\begin{eqnarray}
N_{\textrm{pair}}\left(r\right) &=& S\left(r\right) \frac{V}{R^{3}},
\end{eqnarray}
where
\begin{eqnarray}
S\left(r\right) &=& \frac{2\pi r^{2}dr}{R^{3}}
\end{eqnarray}
is the number of pair sites for one lattice point. When we focus on a certain particle $i$, number of vacant site in distance between $r$ and $r+dr$ is
\begin{eqnarray}
N_{i}\left(r\right) = \frac{4\pi r^{2}dr}{R^{3}} - n_{i}\left(r\right).
\end{eqnarray}
Using above equations and definitions, structural parameter for pair figure in distance $r$ used in the theory can be given by
\begin{eqnarray}
\label{eq:xi}
\xi\left(r\right) = \cfrac{-\rho V \overline{N\left(r\right)} + \left\{ S\left(r\right)\cfrac{V}{R^{3}}-\rho V \overline{N\left(r\right)} \right\} }{S\left(r\right)\cfrac{V}{R^{3}}},
\end{eqnarray}
where $\overline{N\left(r\right)}$ denotes average of $N_{i}\left(r\right)$ over all particles. Using Eqs.~(\ref{eq:g})-(\ref{eq:xi}), we can express radial distribution function by $\xi\left(r\right)$:
\begin{eqnarray}
\label{eq:gr}
g\left(r\right) = \frac{1}{4x^{2}}\xi\left(r\right) + \left(\frac{1}{x} - \frac{1}{4x^{2}}  \right).
\end{eqnarray}
From Eq.~(\ref{eq:gr}), it appears that $g\left(r\right)$ depends on artificially introduced length $R$, but below we see that final representation for {\textit{expectation value}} of $g\left(r\right)$ is independent of $R$. It has been shown\cite{sqs} that pair structural parameter have the characteristic of
\begin{eqnarray}
\left<\xi\left(r\right)\right> = \left(2x-1\right)^{2},
\end{eqnarray}
where $\left<\quad\right>$ denotes average over all possible microscopic states. 
Using the fact from the theory\cite{lsi,emrs,rmx} that density of microscopic states in terms of $\xi\left(r\right)$ is given by normal distribution function, 
we can confirm for average that 
\begin{eqnarray}
\label{eq:lin}
\left<\xi\left(r\right)\right> = 4x^{2}\left<g\left(r\right)\right> - 4x + 1 = \left(2x-1\right)^{2},
\end{eqnarray}
since by definition of radial distribution, $\left<g\left(r\right)\right> = 1$, and for standard deviation, 
\begin{eqnarray}
\label{eq:sd}
\left<g\left(r\right)\right>_{2} &=& \frac{1}{4x^{2}}\left<\xi\left(r\right)\right>_{\textrm{sd}} \nonumber \\
&=&\frac{1}{4x^{2}}\cfrac{1}{\sqrt{\cfrac{V}{R^{3}}S\left(r\right)}}\left(-4x^{2}+4x\right) \nonumber \\
&\simeq& \frac{1}{\rho\sqrt{2\pi r^{2}drV}}.
\end{eqnarray}
To derive the last equation of 
Eq.~(\ref{eq:sd}), we consider the limit of $R\to 0$. Consider that from Eq.~(\ref{eq:lin}), radial distribution function $g\left(r\right)$ is certainly a linear transformation of $\xi\left(r\right)$, expectation value of $g\left(r\right)$ should have the same form as that of $\xi\left(r\right)$ given in the theory. This is because such linear transformation holds multivariate gaussian distribution for density of microscopic states on configuration space, which is required to apply the theory to liquid phase. Using these facts, we get
\begin{eqnarray}
\label{eq:gz0}
\left<g\left(r\right)\right>_{Z} \simeq \left<g\left(r\right)\right>_1 - \sqrt{\frac{\pi}{2}}\frac{1}{k_{\textrm{B}}T}\left<g\left(r\right)\right>_{2}\left\{U_{r}^{+} - U_0  \right\},
\end{eqnarray}
where factor of $\sqrt{2\pi}$ comes from the integration of gaussian from $-\infty$ to $\infty$ in potential energy space.
Here, $U_0$ is the potential energy of mean state that satisfies $g\left(r\right)=\Braket{g\left(r\right)}_1$ for any given $r$.  
$U_{r}^{+}$ is the potential energy of projection state that has $\left<g\left(r'\right)\right>_{r}^{\left(+\right)}$ for any given $r'$, where $\left<\quad\right>_{r}^{\left(+\right)}$ denote average over all microscopic states under the condition of $g\left(r\right) \ge \Braket{g\left(r\right)}_1$. Substituting Eq.~(\ref{eq:sd}) into Eq.~(\ref{eq:gz0}), we obtain the final form of radial distribution function:
\begin{eqnarray}
\label{eq:gfin}
\left<g\left(r\right)\right>_{Z} \simeq \Braket{g\left(r\right)}_1 - \frac{U_{r}^{+} - U_0}{2k_{\textrm{B}}T\rho\sqrt{r^{2}drV}}.
\end{eqnarray}
Eq.~(\ref{eq:gfin}) exhibits general trends for radial distribution in liquid phase, where (i) at high temperature limit ($T\to\infty$), $g\left(r\right)$ approaches to 1 for all $r$ due to dominant contribution of entropy and (ii) for large $r$, $g\left(r\right)$ also approaches to 1, although actual magnitude of $g\left(r\right)$ is still dominated by the numerator of $U_{r}^{+} - U_0$ depending on system.
We should emphasize here that atomic structures with energy of $U_{r}^{+}$ and $U_0$ are independent of both temperature and atomic species, since partial average $\left<g\left(r'\right)\right>_{r}^{\left(+\right)}$ and average $\left<g\left(r\right)\right>$ for microscopic states are clearly independent of temperature and atomic species.

\section{Results and Discussions}
\begin{figure}[h]
\begin{center}
\includegraphics[width=0.8\linewidth]
{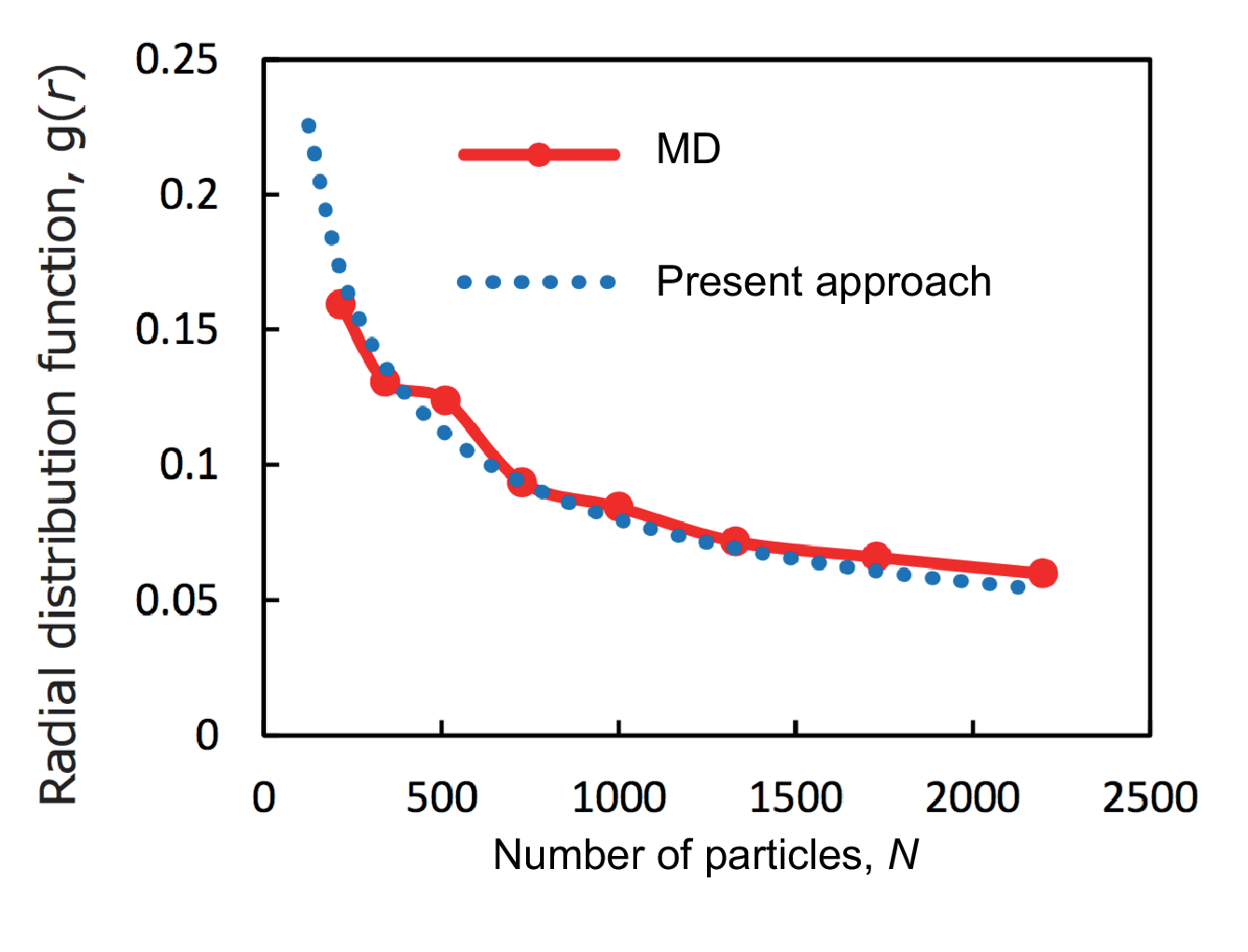}
\caption{Standard deviation of RDF, $\Braket{g\left(r\right)}_2$ at $r=8.5$ \AA, predicted by Eq.~(\ref{eq:sd}) based on our theoretical approach and by MD simulation, as a function of number of particles in the system.}
\label{fig:sd}
\end{center}
\end{figure}

To confirm the applicability of the derived expressions, we compare the RDF predicted by the present approach with that by molecular dynamics (MD) simulations.
Calculation condition of the MD is that we consider$NVT$ ensemble, and employing verlet method to satisfy simplectic conditions, and ......
We employ Lennard-Jones (LJ) pairwise potential for the interaction between constituents, ...
For the condition under non-interacting system, we perform well-type potential where effective radius of particle corresponding to the wall of the well. 
We set the effective radius of constituent particles is HOGE \AA, which is realized by LJ or well-type potentials. 
In order to demonstrate the validity that $g\left(r\right)$ can be a natural linear transformation from GIM, we first compare $\Braket{g\left(r\right)}_2$ given by Eq.~(\ref{eq:sd}) with that by MD simulation  for the chosen distance, $r=8.5$ \AA. 
Figure~\ref{fig:sd} shows the resultant $\Braket{g\left(r\right)}_2$ as a function of number of particles in the system. 
We can clearly see that when the number of particles increases, $\Braket{g\left(r\right)}_2$ given by the present approach successfully agrees with that by MD simulation, indicating the validity of Eq.~(\ref{eq:sd}) as a result of linear transformation from GIM coordination. 

Using the same conditions, we next determine the microscopic structures for mean and projection state based on the MD simulation with non-interacting system. 
Figure~\ref{fig:ms} shows the resultant RDF for mean and projection state, where the sharp peak found for projection state is at $r=8.5$ \AA, which is the same 
distance $r$ that we now focus on: The main difference in pair correlations between mean and projection state is at the considered distance, which is the 
similar tendency found for the crystalline solids, while for higher-body (e.g., 3-body or 4-body) correlations, this does not always hold true. 

\begin{figure}[h]
\begin{center}
\includegraphics[width=0.74\linewidth]
{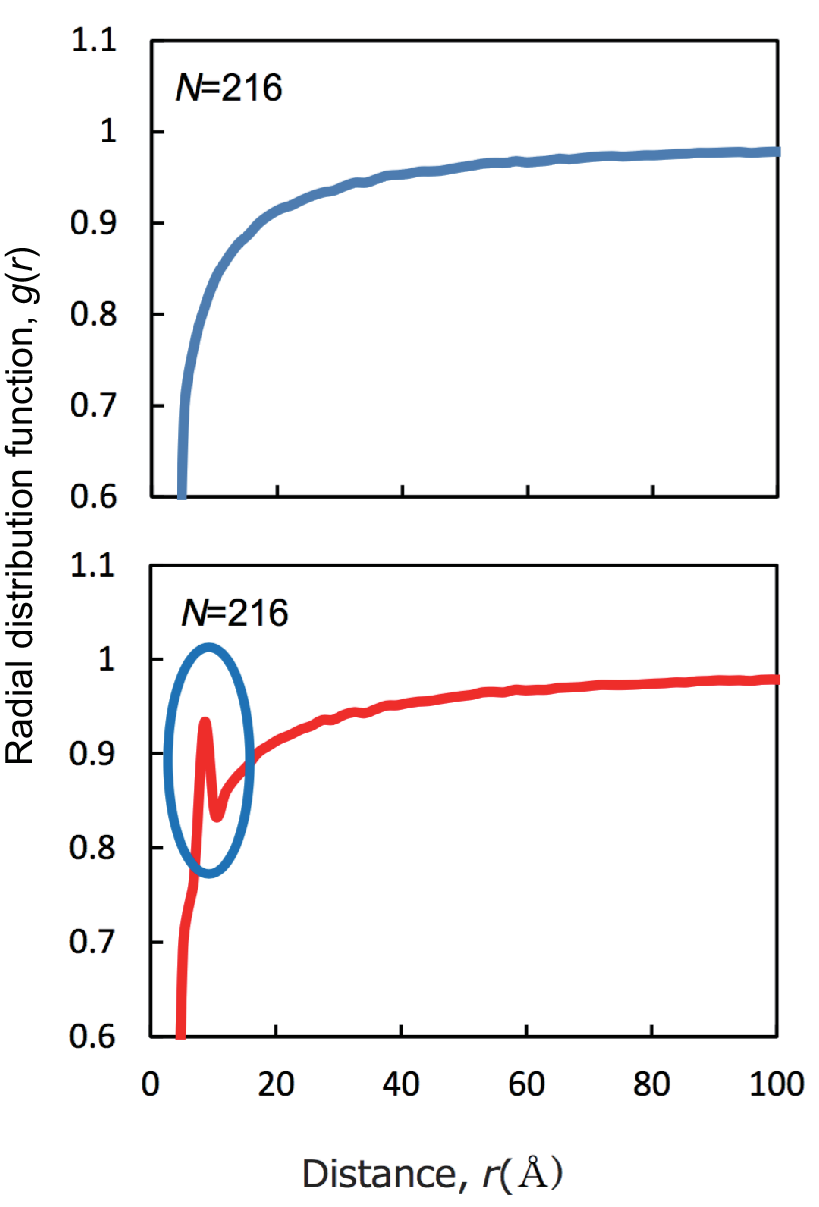}
\caption{Simulated structure of mean and projection states in terms of RDF at $N=216$ and $r=8.5$ \AA. }
\label{fig:mps}
\end{center}
\end{figure}

\begin{figure}[h]
\begin{center}
\includegraphics[width=0.8\linewidth]
{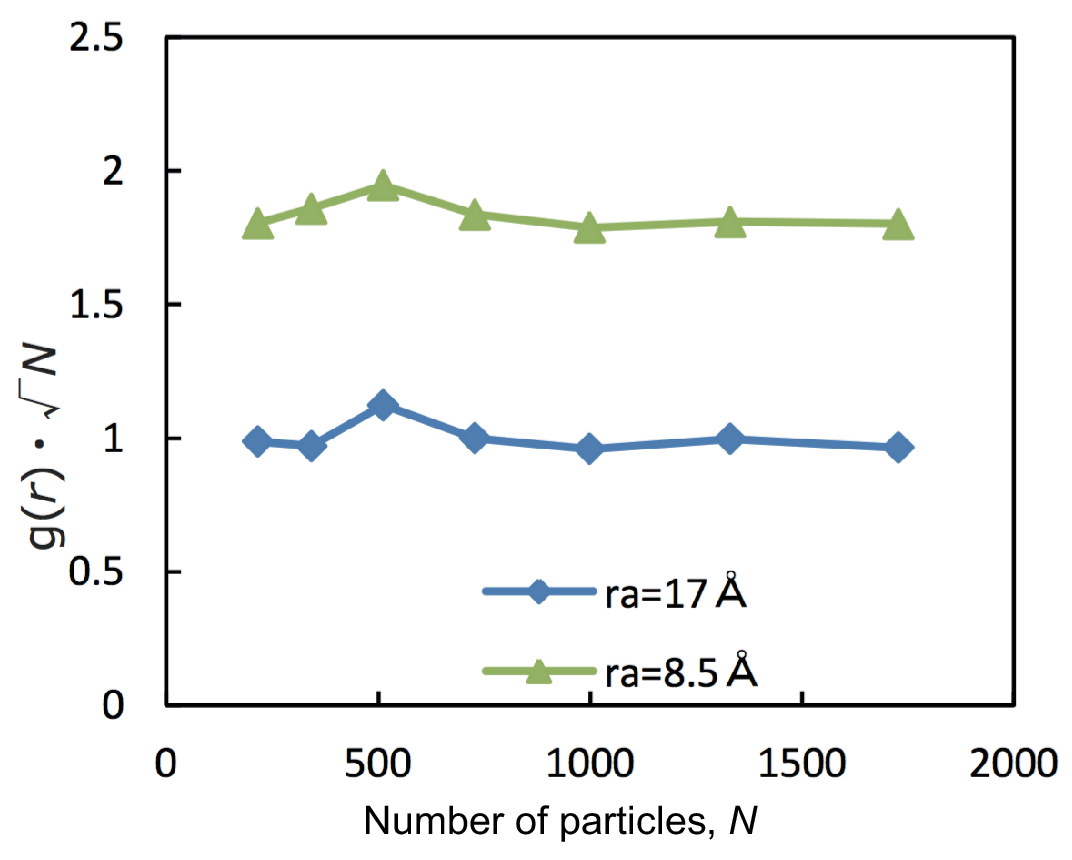}
\caption{Product of $\sqrt{N}$ by RDF of projection state for $r=8.5$ and $r=17$ \AA measured from that of mean state at the same distance, as a function of number of particles, $N$.}
\label{fig:dgr}
\end{center}
\end{figure}

In order to quantitatively confirm the difference in RDF between mean and projection states at the focused distance, we show in Fig.~\ref{fig:dgr} the product of $\sqrt{N}$ by RDF of projection state for $r=8.5$ and 
$r=17$ \AA measured from that of mean state at the same distance, as a function of number of particles, $N$. We can clearly see that the difference in RDF at the focused distance is proportional to $\sqrt{N}$, which is the same tendency as the crystalline solids. 
This certainly indicates that the predicted RDF given by Eq.~(\ref{eq:gfin}) does not in principle depend on the choice of at which number of atoms projection state is constructed: 
Since from Fig.~\ref{fig:dgr}, we can see that numerator of Eq.~(\ref{eq:gfin}), $U_r^+ - U_0$, is proportional to $\sqrt{N}$ (because energy is \textit{extensive}), while denominator also is 
proportional to $\sqrt{N}$ since $V\propto N$ under constant $\rho$. Consequently, $\Braket{g\left(r\right)}_Z$ in Eq.~(\ref{eq:gfin}) is independent of number of atoms for projection state. 
Note that difference in magnitude of $\Delta g\left(r\right)\cdot \sqrt{N}$ between $r=8.5$ and $r=17$ \AA$ $ certainly comes from the difference in volume to estimate $g\left(r\right)$ under constant $dr$.

With these considerations, we finally confirm whether the microscopic structure of projection state, shown in Fig.~\ref{fig:mps}, can be numerically constructed by a single state. 
We perform Monte Carlo (MC) simulation to optimize the RDF differences between ideal and practical systems based on simulated annealing algorism by randomly moving selected 
particles to vacant position. 

\begin{figure}[h]
\begin{center}
\includegraphics[width=0.8\linewidth]
{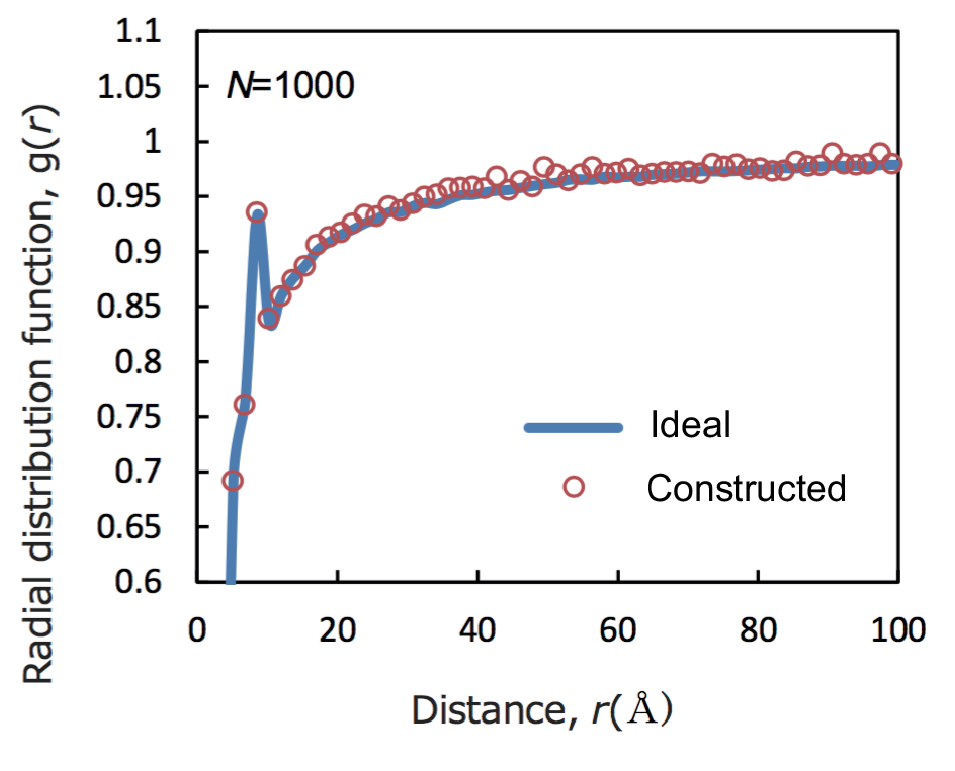}
\caption{RDF of projections state at $N=216$ for ideal and constructed structures. Note that constructed structures includes 1000 particles. }
\label{fig:ps}
\end{center}
\end{figure}

Figure~\ref{fig:ps} shows the comparison of RDF between ideal and constructed structures, which indicates that projection state for non-solid matter can be numerically constructed 
within a finite number of particles. 

 \section*{Acknowledgement}
This work was supported by a Grant-in-Aid for Scientific Research (16K06704) from the MEXT of Japan, Research Grant from Hitachi Metals$\cdot$Materials Science Foundation, and Advanced Low Carbon Technology Research and Development Program of the Japan Science and Technology Agency (JST).

\end{document}